\documentclass[conference]{IEEEtran}
\IEEEoverridecommandlockouts
\usepackage{booktabs}
\usepackage{cite}
\usepackage{amsmath,amssymb,amsfonts}
\usepackage{algorithmic}
\usepackage{graphicx}
\usepackage{textcomp}
\usepackage{xcolor}
\usepackage{mathtools}
\definecolor{commentred}{rgb}{0.6,0,0} 
\usepackage[linesnumbered,ruled,vlined]{algorithm2e}
\SetCommentSty{mycommfont}

\usepackage{pgfplots}
\pgfdeclarelayer{background}
\pgfdeclarelayer{foreground}
\pgfsetlayers{background,main,foreground}
\pgfplotsset{compat=newest}
\usepackage{pgfplotstable}
\usepackage{tikz}
\usetikzlibrary{chains,
                fit,
                positioning,
                shapes,
                shadows,
                trees,
                patterns,
                matrix}
\definecolor{darkred}{rgb}{0.55, 0.0, 0.0}
\definecolor{darkgoldenrod}{rgb}{0.72, 0.53, 0.04}
\definecolor{darkkhaki}{rgb}{0.74, 0.72, 0.42}
\definecolor{navyblue}{rgb}{0.0, 0.0, 0.5}
\definecolor{moonstoneblue}{rgb}{0.45, 0.66, 0.76}
\definecolor{ashgrey}{rgb}{0.7, 0.75, 0.71}
\definecolor{blond}{rgb}{0.98, 0.94, 0.75}
\definecolor{burntumber}{rgb}{0.54, 0.2, 0.14}
\definecolor{darkcyan}{rgb}{0.0, 0.55, 0.55}

\definecolor{color1}{rgb}{0.215686,0.494118,0.721569}
\definecolor{color2}{rgb}{0.894118,0.101961,0.109804}
\definecolor{color3}{rgb}{0.301961,0.686275,0.290196}

\usepgfplotslibrary{colorbrewer}
\pgfplotsset{cycle list/Set1}
\pgfplotsset{cycle list/Set2}
\pgfplotsset{cycle list/Set3}
\pgfplotsset{cycle list/Dark2}
\usepgfplotslibrary{groupplots}

\IEEEeqnarraydefcolsep{0}{\leftmargini} 

\def\BibTeX{{\rm B\kern-.05em{\sc i\kern-.025em b}\kern-.08em
    T\kern-.1667em\lower.7ex\hbox{E}\kern-.125emX}}
    
\usepackage[flushleft]{threeparttable}

\makeatletter
    \newcommand{\linebreakand}{%
      \end{@IEEEauthorhalign}
      \hfill\mbox{}\par
      \mbox{}\hfill\begin{@IEEEauthorhalign}
    }
\makeatother

\begin{document}

\newcommand{\tb}{\textbf}

\title{ParDen: Surrogate Assisted Hyper-Parameter Optimisation for Portfolio Selection}


\author{
\IEEEauthorblockN{1\textsuperscript{st} T. L van Zyl\IEEEauthorrefmark{1},
2\textsuperscript{nd} M Woolway\IEEEauthorrefmark{2},
3\textsuperscript{rd} A Paskaramoorthy\IEEEauthorrefmark{3}}
\IEEEauthorblockA{\IEEEauthorrefmark{1}\textit{Institute for Intelligent Systems}\\
\textit{University of Johannesburg}\\
Johannesburg, South Africa\\
Email: tvanzyl@uj.ac.za}
\IEEEauthorblockA{\IEEEauthorrefmark{2}\textit{Faculty of Engineering and the Built Environment}\\
\textit{University of Johannesburg}\\
Johannesburg, South Africa\\
Email: matt.woolway@gmail.com}
\IEEEauthorblockA{\IEEEauthorrefmark{3}\textit{School of Computer Science and Applied Mathematics}\\
\textit{University of the Witwatersrand}\\
Johannesburg, South Africa\\
Email: andrew.paskaramoorthy@wits.ac.za}
}

\maketitle

\begin{abstract}
Portfolio optimisation is a multi-objective optimisation problem (MOP), where an investor aims to optimise the conflicting criteria of maximising a portfolio's expected return whilst minimising its risk and other costs. However, selecting a portfolio is a computationally expensive problem because of the cost associated with performing multiple evaluations on test data (``backtesting'') rather than solving the convex optimisation problem itself. In this research, we present ParDen, an algorithm for the inclusion of any discriminative or generative machine learning model as a surrogate to mitigate the computationally expensive backtest procedure. In addition, we compare the performance of alternative metaheuristic algorithms: NSGA-II, R-NSGA-II, NSGA-III, R-NSGA-III, U-NSGA-III, MO-CMA-ES, and COMO-CMA-ES. We measure performance using multi-objective performance indicators, including Generational Distance Plus, Inverted Generational Distance Plus and Hypervolume. We also consider meta-indicators, Success Rate and Average Executions to Success Rate, of the Hypervolume to provide more insight into the quality of solutions. Our results show that ParDen can reduce the number of evaluations required by almost a third while obtaining an improved Pareto front over the state-of-the-art for the problem of portfolio selection.
\end{abstract}


\section{Introduction}\label{sec:introduction}
Portfolio optimisation is a multi-objective optimisation problem (MOP), where an investor aims to optimise the conflicting criteria of maximising a portfolio's expected return whilst minimising its risk, and other costs \cite{boyd2017multi}. The original mean-variance optimisation problem is specified by a bi-objective criterion, where risk is measured by the variance of the portfolio's return \cite{markowitz1952}. These objectives are combined linearly into a single optimisation problem:
\begin{eqnarray}
    &\text{maximise} \quad &\mathbf{w}'\mathbf{\mu} - \frac{\gamma}{2}\mathbf{w}'\Sigma \mathbf{w} \\
    &\text{subject to} \quad &\mathbf{w}'\mathbf{1} = 1
\end{eqnarray}
where $\mathbf{w}$ is the vector of portfolio weights, $\mathbf{\mu}$ is the vector of expected returns, $\Sigma$ is the covariance matrix of returns, and $\gamma$ is known as the \textit{risk-aversion} parameter, which is used to trade-off the conflicting objectives. The resulting problem is a quadratic program that has an analytical solution. A particular choice of the risk aversion parameter yields a particular optimal solution. By varying the risk-aversion parameter and solving the optimisation problem, we can uncover the set of non-dominated solutions known as the Pareto frontier or, in the finance jargon, the efficient frontier.

In practice, portfolio selection problems are specified more extensively by incorporating a wide range of real-world considerations \cite{boyd2017multi}. For example, the optimisation problem can include additional objectives for trading costs and holding costs, as well as a leverage constraint:
\begin{eqnarray}
    \text{maximise} \quad &&\mathbf{w}'\mathbf{\mu} - \frac{\gamma}{2}\mathbf{w}'\Sigma \mathbf{w} - \gamma_{t}\phi^\text{trade}\left(\mathbf{w} - \mathbf{w}_0\right) \nonumber\\
    && \qquad - \gamma_h \phi^\text{hold}\left(\mathbf{w}\right) \\
    \text{subject to} \quad &&\lVert\mathbf{w}\rVert_1 \leq L^\text{max}
\end{eqnarray}

Here, $\mathbf{w}_0$ is the initial portfolio, the difference $\mathbf{w} - \mathbf{w}_0$ denotes the trades required to attain portfolio $\mathbf{w}$, $\phi^\text{trade}$ and $\phi^\text{hold}\left(\mathbf{w}\right)$ are the trading and holding cost functions with corresponding trade-off parameters $\gamma_t$ and $\gamma_h$, and maximum leverage $L^{\text{max}}$. 

Additional considerations can feature in the problem as hard constraints or otherwise can be included as additional objectives (i.e. soft constraints), depending on an investor's particular needs. In the latter case, these additional objectives are added to the mean-variance objective function and have associated trade-off parameters. The inclusion of hard and soft constraints implies that the mean-variance optimisation is no longer analytically tractable. 

The use of metaheuristic algorithms for finding solutions to portfolio-optimisation problems has been covered extensively in the literature \cite{art:doering2019metaheuristics,art:loukeris2009numerical,art:fernandez2015hybrid}. To motivate new algorithms, the convex sub-problem is contrived with appropriate constraints to make it non-convex. Nonetheless, most real-world considerations can be included in a convex specification of the optimisation problem and hence can be easily solved using Disciplined Convex Programming (DCP) \cite{cvx,grant2006disciplined}. Here, a convex solver is used to produce an exact solution to a particular set of trade-off parameters. The Pareto frontier can be uncovered by repeatedly solving the problem for different choices of the trade-off parameters.  

Motivated by hyper-parameter search, a standard practice in machine learning, Boyd \textit{et al.}~\cite{boyd2017multi} considered the set of trade-off parameters as the hyper-parameters and the portfolio weights as the parameters in a hybrid approach. First, they specify the portfolio weights optimisation as a single-objective convex problem and use DCP to solve the corresponding exact solution. They then used a grid search over the hyper-parameters to uncover the subset of non-dominated solutions making up the Pareto-frontier. However, without prior knowledge of suitable trade-off parameters to evaluate, this process is typically computationally inefficient as many of the chosen trade-off parameters will result in sub-optimal solutions.

Nystrup~\cite{nystrup2020hyperparameter} extends upon the work of Boyd \textit{et al.}~\cite{boyd2017multi} by using a metaheuristic algorithm to generate the trade-off parameter set. They investigate using a single metaheuristic algorithm, the Multi-Objective Covariance-Matrix Adaptation Evolution Strategy (MO-CMA-ES), to generate new trade-off parameters to evaluate. By combining a metaheuristic and exact-solution approach, they can generate exact solutions on the frontier using far fewer evaluations than the previous grid search procedure. Other researchers have also considered the hybrid approach with similar success to both Boyd \textit{et al.}~\cite{boyd2017multi} and Nsytrup~\cite{nystrup2020hyperparameter}, showing its utility over a purely metaheuristic approach when the sub-problem can be solved using convex optimisation~\cite{branke2009portfolio,ruiz2010hybrid,ruiz2015memetic,baykasouglu2015grasp,qi2017hybrid}.

In these scenarios, the computational intractability of evaluating all possible parameter combinations arises from the cost associated with performing multiple evaluations on test data (``backtesting'') rather than from solving the convex optimisation problem itself. In contrast, multi-objective algorithms can solve the Pareto frontier in a single stage by simultaneously solving both the trade-off parameters and the corresponding solutions. However, this gain in computational efficiency is at the expense of finding approximate solutions. 

In this research, we adopt the hybrid approach of Boyd \textit{et al.} \cite{boyd2017multi} and consider extensions to reduce the number of evaluations of the backtest. We first extend the research of Nystrup \cite{nystrup2020hyperparameter} by comparing his usage of MO-CMA-ES against the performance of the alternative metaheuristic algorithms: NSGA-II, R-NSGA-II, NSGA-III, R-NSGA-III U-NSGA-III and COMO-CMA-ES. We then build on this by considering the inclusion of a surrogate model to mitigate the computationally expensive backtest procedure. Our results include both the original grid-search of Boyd \textit{et al.}~\cite{boyd2017multi} as well as a random search as baselines for comparison. In order to allow for quantitative comparison, we introduce several performance indicators, including Generational Distance Plus (GD+), Inverted Generational Distance Plus (IGD+) and Hypervolume (HV). We then consider several meta-indicators of the HV to provide more insight into the quality of solutions. In particular, we consider the Success Rate (SR), the Average Evaluations to SR (AESR) and Average Generations to SR (AGSR)~\cite{book:eiben2015}.

\section{Background}

\subsection{Metaheuristic Multi-Objective Optimisation Methods}

There exists a significant number of metaheuristic MOP algorithms in the literature~\cite{art:liu2020multi}. Amongst these, certain algorithms have been shown to be successful in applications across several different domains~\cite{art:deb2002fast,art:zhang2007moea,art:igel2007covariance,art:helbig2014population,art:alsattar2020mogsabat}. Our objective is to apply various representative metaheuristic algorithms and demonstrate the utility of a surrogate assisted metaheuristic optimisation approach to solving portfolio optimisation problems of the nature described in this paper. To this end, we implement the metaheuristic MOP methods described below.

We use Latin Hyper-cube Sampling for random selection for all our algorithms, we use real Uniform Crossover (UX) for crossover, and we use real Polynomial Mutation (PM) for mutation. For real UX, we set the probability of crossover to 90\%. For PM, we set the probability of mutation to 20\%. For all techniques, we use a population size of 60 and offspring size of 30. For all techniques, we give a total evaluation budget of $510$.

\subsubsection{MO-CMA-ES}

In multi-objective CMAES (MO-CMA-ES), a population of individuals that adapt their search strategy as in the elitist CMA-ES is maintained. The elites are subjected to multi-objective selection pressure. The selection pressure originates from non-dominated sorting using the crowding-distance first and then hypervolume as the second criterion~\cite{blank2020pymoo}. We set sigma – the initial step size of the complete system - to $0.1$. Other parameters to CMA-ES are left as their defaults.

\subsubsection{NSGA-II}

The algorithm follows the form of a genetic algorithm with modified mating and survival.  The individuals are chosen Pareto front-wise first. There will arise a situation where a front needs to be split since not all individuals can survive. When splitting the front, solutions are selected based on a Manhattan crowding distance in the objective space. However, we want to hold onto the extreme points from each generation, and as a result, they are assigned a crowding distance of infinity. Furthermore, to increase selection pressure, we use a binary tournament mating selection. Each individual is first compared by rank and then crowding distance~\cite{blank2020pymoo}. 

\subsubsection{R-NSGA-II}

The algorithm follows NSGA-II with modified survival selection. However, unlike NSGA-II, in splitting the front, solutions are selected based on rank. This rank is calculated based on the euclidean distance to a set of supplied reference points. The solution closest to a reference point is assigned a rank of one, with each point taking its best rank. 
After each reference point has selected the solution with the best rank for survival, all solutions within the epsilon distance of surviving solutions are ``cleared'', meaning they can not be selected for survival until all fronts have been processed. If more solutions are still needed to be selected, the cleared points are considered. The free parameter epsilon has the effect that a smaller value results in a tighter set of solutions~\cite{blank2020pymoo}. For our work, we choose an epsilon of 0.1 by manual tuning.

\subsubsection{NSGA-III}

The algorithm is based on reference directions instead of reference points. For survival, after non-dominated sorting, the algorithm has a modified mechanism for dealing with the front splitting. The algorithm fills up the underrepresented reference direction first. If the reference direction does not have any solution assigned, then the solution with the smallest perpendicular distance in the normalised objective space is assigned as the survivor. In the case that additional solutions for a reference direction are required, these are assigned randomly. Consequently, when this algorithm converges, each reference direction attempts to find a suitable non-dominated solution. We use Riesz s-Energy to generate reference directions~\cite{blank2020pymoo}.

\subsubsection{U-NSGA-III} 

It has previously been shown that tournament selection performs better than random selection. Unlike NSGA-III, which selects parents randomly for mating, this algorithm uses tournament pressure as an improvement~\cite{blank2020pymoo}.  We use Riesz s-Energy to generate reference directions~\cite{blank2020pymoo}.

\subsubsection{R-NSGA-III} 

The algorithm follows the general NSGA-III procedure with a modified survival selection operator. First, non-dominated sorting is done as in NSGA-III. Solutions are associated with aspiration points based on perpendicular distance, then solutions from the underrepresented reference direction are chosen. For this reason, when this algorithm converges, each reference line seeks to find a good representative non-dominated solution~\cite{blank2020pymoo}.

\subsubsection{COMO-CMA-ES}

The algorithm is a multi-objective version of the Covariance Matrix Adaptation Evolution Strategy (CMA-ES) single objective optimiser. Each single-objective CMA-ES optimises an indicator function given $p-1$ fixed solutions. Dominated solutions try to minimise their distance to the empirical Pareto front defined by these $p-1$ solutions. Other parameters to CMA-ES are left as their defaults~\cite{blank2020pymoo}.

\subsection{Surrogate Assisted Metaheuristic Multi-Objective Optimisation}

Although surrogate methods have been extensively applied within a Bayesian framework aimed at single objective optimisation~\cite{art:ben2017universal,art:zhou2006combining,art:wan2005simulation}, the applications of surrogate assisted methods within a multi-objective setting are substantially less common. This is especially so for data-driven metaheuristic multi-objective surrogate assisted optimisation~\cite{art:wang2018random,art:yang2019offline,art:chugh2017data,art:stander2020data}. To this end, we present our algorithm, ParDen, which is a Pareto driven multi-objective optimisation. ParDen is sufficiently general that it can be applied together with most metaheuristic multi-objective algorithms. It also allows us to move beyond using generative models as surrogates, allowing for the future integration of discriminative models such as deep neural networks and recurrent neural networks.

ParDen relies on the use of any supervised learning model to act as a surrogate drop-in replacement $\hat{f}(\cdot)$ to evaluate a computationally resource intensive real-world experiment or simulation $f_e(\cdot)$. In this research, the ``backtesting'' procedure, which runs for several minutes per a set of trade-off parameters, is the simulation under consideration.


The core idea of ParDen is to limit our evaluation of candidates using the simulation $f_e(\cdot)$ to only those that have a high probability of advancing the Pareto front (the non-dominated solution set $\mathcal{P}$). The likelihood that a candidate would advance the Pareto front is determined by two things: i) whether the surrogate model predicts that the candidate advances the Pareto front and ii) the extent to which our surrogate is correct. These two criteria allow us to balance exploration and exploitation of the surrogate. In particular, we attempt to estimate the probability that the surrogate is correct, and we use this probability as a threshold within a rejection sampling scheme to select candidates.

In order to determine the extent to which our surrogate is correct, we require a mechanism for assessing the probability $P[(x_i,y_i) \in \mathcal{P} | \hat{f}]$ that a pretender $(x_i, y_i)$ will be in the current non-dominated set $\mathcal{P}$ given the surrogate's prediction $\hat{y}_i = \hat{f}(\cdot)$. We employ $k$-fold cross-validation $\operatorname{CV}_k(\cdot)$ together with the non-dominating rank~\cite{book:selvi2018application} of each candidate to estimate this probability as a non-dominated score ($\operatorname{NDScore}$). 

To calculate a $\operatorname{NDScore}$ we fit surrogates $\hat{f}_{_{X_T,Y_T}}(\cdot)$ using the training splits $(X_T,Y_T)$ with the multi-objective $Y_T$ as the target. For each point in the validation split $(X_V,Y_V)$, we assign its non-dominating rank in the split as its class label $r$. We use the surrogates to estimate $\hat{Y}_V=\hat{f}_{_{X_T,Y_T}}(X_V)$, the objective values of the validation split. We then assign the non-dominating rank of the estimated values $\hat{Y}_V$ as the predicted class labels $\hat{r}$. We can use any classification metric $\mathcal{E}(\hat{r}, r)$ constrained to sum to one, to compare $r$ with $\hat{r}$: 
\begin{equation}
\operatorname{NDScore}(\cdot) = \operatorname*{aggregation}_{\substack{(X_T,Y_T),(X_V,Y_V)\\ \in \operatorname{CV}_k(X)}} \mathcal{E}\left(\hat{r}, r\right) \\
\end{equation}
with 
\begin{equation}
\hat{r} = \lceil Y_V \rceil^{\uparrow} \quad \mathrm{and} \quad
      r = \lceil\hat{f}_{_{X_T,Y_T}}(X_V) \rceil^{\uparrow}
\end{equation}
where $\lceil\cdot\rceil^\uparrow$ assigns the non-dominating rank to each vector. Since we are using $k$-fold cross-validation to estimate this value and we are interested in the worst case, we use the minimum across all $k$ runs as our aggregation. We note that accuracy could be replaced with a more balanced metric such as $F_1$-Score and that using the mean instead of minimum might be a better aggregation over $CV_k$. We did not investigate this further.


\begin{algorithm}[htbp]
\caption{ParDen Algorithm}
\SetKwInput{KwInput}{Input}                
\SetKwInput{KwOutput}{Output}              
\DontPrintSemicolon
\SetNoFillComment 
  
  \KwInput{\textit{Metaheuristic} ($M$), \textit{Terminating Criterion} ($TC$), \textit{Simulation} ($f_e$), \textit{Surrogate Class} ($\mathcal{H}$), \textit{Loss} ($L$) }
  \KwOutput{\textit{Approximate Optimal Frontier ($\mathcal{P}$)}}
  \KwData{\textit{Daily Trading Data} ($D$)}

  \SetKwFunction{FMain}{main}
  \SetKwFunction{Init}{init}
  \SetKwFunction{FSub}{Sub}

  \SetKwProg{Fn}{Initialise}{:}{}
  \Fn{\Init{$M, f_e, \ldots$}}{
        $X_W \sim M$                \tcc*[r]{Generate warm start candidates}
        $Y_W=f_e(X_W,D)$              \tcc*[r]{Actual simulated fitness}
        $\mathcal{G}=\{(X_W,Y_W)\}$ \tcc*[r]{Set ground-truth}
        $\mathcal{P} = \lceil \mathcal{G} \rceil$ \tcc*[r]{Compute non-dominated set}   
        $\operatorname{NDScore} = 0$  \tcc*[r]{Set non-dominated score}
        \tcc*[r]{Train surrogate on $\mathcal{G}$ with loss $L$}
        $\hat{f} = \operatorname{argmin}_{h \in \mathcal{H}} \mathbb{E}[L(h,\mathcal{G})]$ \;
        \KwRet $\hat{f}, \mathcal{G}, \operatorname{NDScore}, \mathcal{P}$\;
  }
  
  \SetKwProg{Fn}{Function}{:}{\KwRet}
  \Fn{\FMain{$M, f_e, \ldots$}}{
        $X_C \sim M$                \tcc*[r]{Generate new candidates}
        \tcc*[r]{Estimate candidates' fitness with surrogate}
        $\hat{Y}_C=\hat{f}(X_C, D)$\;
        \tcc*[r]{Join candidates to non-dominated set}
        $\mathcal{P}_C = \lceil \{\mathcal{C} \cup \mathcal{P}\} \rceil$  where $\mathcal{C}=\{(X_C,\hat{Y}_C)\}$\; 
        \tcc*[r]{Pretenders are non-dominated candidates}
        ($X_S, \hat{Y}_S) = \{\mathcal{P}_C \cup \mathcal{C}\}$\;
        \tcc*[r]{Acceptance sampling with $\operatorname{NDScore}$ as threshold to add additional pretenders}
        $X_S \quad\operatorname{append}\quad X_C$ if $\operatorname{NDScore} \le x \sim [0,1]$\;
        $Y_S=f_e(X_S,D)$      \tcc*[r]{Actual pretenders' fitness}
        $\mathcal{G} = \lbrace\mathcal{G}\cup (X_S,Y_S) \rbrace$ \tcc*[r]{Add to ground-truth}
        $\mathcal{P} = \lceil \mathcal{G} \rceil$ \tcc*[r]{Update non-dominated set}   
        \tcc*[r]{Update non-dominated score}
        $\operatorname{NDScore} = \operatorname{NDScore}(\hat{f}, \mathcal{G})$ \;
       \tcc*[r]{Train new surrogate on $\mathcal{G}$ with loss $L$}
        $\hat{f} = \operatorname*{argmin}_{h \in \mathcal{H}} \mathbb{E}[L(h,\mathcal{G})]$ \;
        \KwRet $\hat{f}, \mathcal{G}, \operatorname{NDScore}, \mathcal{P}$\;
  }

  \texttt{init} ($M, f_e$)\;
  \While{$TC$ = \textit{False}}
   {
   		\texttt{main} ($M, f_e, \ldots$)\;
   		\KwRet $\mathcal{P}$\;
   }\;
   $X \sim I$: sample $X$ from candidate generator $I$\;
   $\lceil \mathcal{X} \rceil$: Return non-dominated set of $\mathcal{X}$\;
\end{algorithm}

\tikzstyle{startstop} = [rectangle, rounded corners, minimum width=2cm, minimum height=1cm,text centered, draw=ashgrey, drop shadow, fill=ashgrey!50,]
\tikzstyle{io} = [trapezium, trapezium left angle=70, trapezium right angle=110, minimum width=2cm, minimum height=1cm, text centered, draw=black, drop shadow]
\tikzstyle{process} = [rectangle, minimum width=4.5cm, minimum height=1cm, text centered, text width=4cm, draw=orange, drop shadow, fill = orange!20]
\tikzstyle{process2} = [rectangle, minimum width=2cm, minimum height=1cm, text centered, text width=2cm, draw=black, drop shadow, fill = white]
\tikzstyle{process3} = [rectangle, minimum width=4.0cm, minimum height=1cm, text centered, text width=4.0cm, draw=orange, drop shadow, fill = orange!20]
\tikzstyle{process4} = [rectangle, minimum width=2.0cm, minimum height=1cm, text centered, text width=2.0cm, draw=orange, drop shadow, fill = orange!20]
\tikzstyle{decision} = [diamond, minimum width=3cm, minimum height=1cm, text centered, draw=darkred, drop shadow, fill = darkred!20]
\tikzstyle{arrow} = [thick,->,>=stealth]
\tikzset{Tank/.style={draw=black,fill=white,thick,rectangle,rounded corners=10pt,minimum width=0.75cm,minimum height=1.5cm,text width=0.75cm,align=center, drop shadow}}
\tikzstyle{surround} = [fill=gray!15,thick,draw=black,rounded corners=2mm]

\tikzstyle{line} = [draw, -stealth]
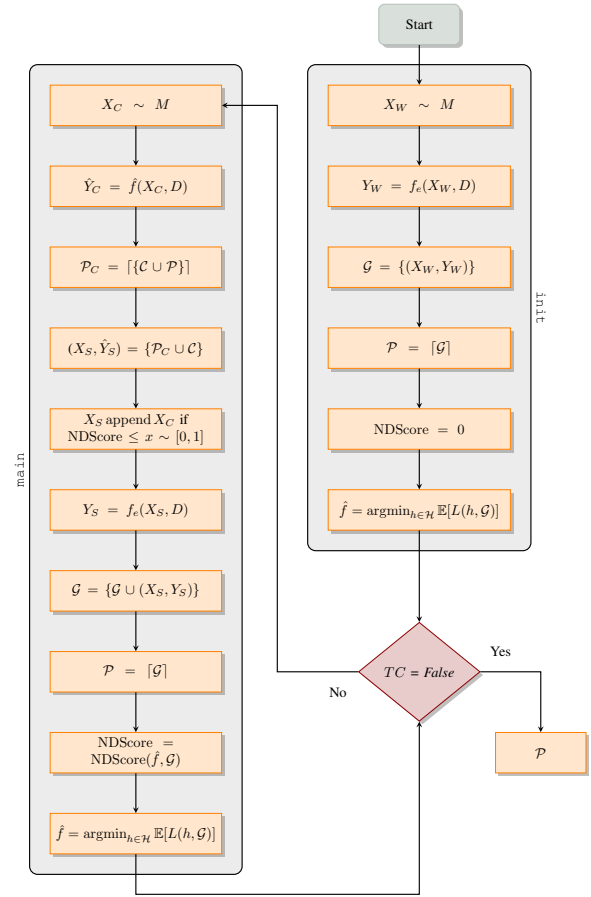
\begin{figure}[htbp]
\centering
\resizebox{0.43\textwidth}{!}{
\begin{tikzpicture}[node distance=1cm,every node/.style={fill=white}, align=center]


\node (start) [startstop,xshift=3cm] {Start};


\node (pro1) [process, below of =start, yshift=-1cm] {$X_W \sim M$};
\node (pro2) [process, below of =pro1, yshift=-1cm] {$Y_W=f_e(X_W,D)$};
\node (pro3) [process, below of =pro2, yshift=-1cm] {$\mathcal{G}=\{(X_W,Y_W)\}$};
\node (pro4) [process, below of =pro3, yshift=-1cm] {$\mathcal{P} = \lceil \mathcal{G} \rceil$};
\node (pro5) [process, below of =pro4, yshift=-1cm] {$\operatorname{NDScore} = 0$};
\node (pro6) [process, below of =pro5, yshift=-1cm] {$\hat{f} = \operatorname{argmin}_{h \in \mathcal{H}} \mathbb{E}[L(h,\mathcal{G})]$};

\begin{pgfonlayer}{background} 
    \node[surround] (background) [inner sep=0.5cm, fit = (pro1) (pro6), label={[rotate=270, anchor=south]right:\texttt{init}}] {};
\end{pgfonlayer}

\node (dec2) [decision, below of=pro5, yshift=-5cm] {$TC$ = \textit{False}};


\node (main1) [process3, left of =pro1, xshift=-6cm] {$X_C \sim M$};
\node (main2) [process3, below of =main1, yshift=-1cm] {$\hat{Y}_C=\hat{f}(X_C, D)$};
\node (main3) [process3, below of =main2, yshift=-1cm] {$\mathcal{P}_C = \lceil \{\mathcal{C} \cup \mathcal{P}\} \rceil$};
\node (main4) [process3, below of =main3, yshift=-1cm] {($X_S, \hat{Y}_S) = \{\mathcal{P}_C \cup \mathcal{C}\}$};
\node (main5) [process3, below of =main4, yshift=-1cm] {$X_S \operatorname{append} X_C$ if $\operatorname{NDScore} \le x \sim [0,1]$};
\node (main6) [process3, below of =main5, yshift=-1cm] {$Y_S=f_e(X_S,D)$};
\node (main7) [process3, below of =main6, yshift=-1cm] {$\mathcal{G} = \lbrace\mathcal{G}\cup (X_S,Y_S) \rbrace$};
\node (main8) [process3, below of =main7, yshift=-1cm] {$\mathcal{P} = \lceil \mathcal{G} \rceil$};
\node (main9) [process3, below of =main8, yshift=-1cm] {$\operatorname{NDScore} = \operatorname{NDScore}(\hat{f}, \mathcal{G})$};
\node (main10) [process3, below of =main9, yshift=-1cm] {$\hat{f} = \operatorname*{argmin}_{h \in \mathcal{H}} \mathbb{E}[L(h,\mathcal{G})]$};
\begin{pgfonlayer}{background} 
    \node[surround] (background2) [inner sep=0.5cm, fit = (main1) (main10), label={[rotate=90, anchor=south]left:\texttt{main}}] {};
\end{pgfonlayer}


\node (out1) [process4, right of =dec2, xshift=2cm, yshift=-2cm] {$\mathcal{P}$};


\draw [arrow] (start) -- (pro1);
\draw [arrow] (pro1) -- (pro2);
\draw [arrow] (pro2) -- (pro3);
\draw [arrow] (pro3) -- (pro4);
\draw [arrow] (pro4) -- (pro5);
\draw [arrow] (pro5) -- (pro6);

\draw [arrow] (pro6) -- (dec2);
\draw [arrow] (dec2.west) node[below, xshift=-0.5cm, yshift=-0.25cm]{No} |- ++ (-2.0cm, -0cm) |- (main1);

\draw [arrow] (main10) |- ++(0, -15mm) -| (dec2.south);

\draw [arrow] (dec2.east) node[below, xshift=0.5cm, yshift=0.75cm]{Yes} -| (out1);

\draw [arrow] (main1) -- (main2);
\draw [arrow] (main2) -- (main3);
\draw [arrow] (main3) -- (main4);
\draw [arrow] (main4) -- (main5);
\draw [arrow] (main5) -- (main6);
\draw [arrow] (main6) -- (main7);
\draw [arrow] (main7) -- (main8);
\draw [arrow] (main8) -- (main9);
\draw [arrow] (main9) -- (main10);
\end{tikzpicture}
}
\caption{Flowchart for the ParDen Algorithm}
\label{fig:parden_framework}
\end{figure}

\section{Methodology}\label{sec:methodology}

In order to demonstrate the efficacy of the ParDen algorithm, we first evaluate several MOP EAs on our problem to see which of them performs best and provide a baseline. We then test ParDen in the two best performing EAs: MO-CMA-ES and NSGA-II. Since we cannot clearly ascertain a single best EA, we instead select the best one with a fixed number of points in the front and one with a dynamic number of points in the front. Finally, we compare all of these results using the following quality metrics on the HV score.

\textbf{Success Rate (SR):} Success is defined as a required quality criterion that needs to be achieved, i.e. a solution within 99\% or 95\% of the known optimal value. The success rate as a metric is defined as the percentage of algorithm executions in which the success criterion is met. 

\textbf{Average Evaluations to a Solution (AES):} Average number of evaluations of the simulation required to reach a solution at a given SR, i.e. the AES at a SR of 99\%, would be the average number of evaluations it took to reach there. The particular runs of the EA which do not meet the SR criterion are ignored.

\textbf{Average Generations to Success Rate (AGSR):} Average number of generations of the simulation required to reach the solution at a given SR, i.e. the AGSR at a SR of 99\%, would be the average number of generations required to reach there.


\section{Results and Discussion}\label{sec:results}

\begin{table}[htbp]
    \begin{threeparttable}
    \caption{Performance Indicators}
    \label{tab:perfermance_indicators}
    \centering
    \begin{tabular*}{\columnwidth}{@{\extracolsep{\fill}} l rrr}
    \toprule
        Method                                      & GD$+$      & IGD$+$     & HV \\
    \toprule
        Random $6510$                               & -          & -          & 1313.0 \\
        Grid Search~\cite{boyd2017multi}            & .5941      &  .5703     & 1259.1 \\
        NSGA-II                                     & .1137      &  .1585     & 1308.4 \\
        R-NSGA-II                                   & \tb{.0259} &  .2443     & 1309.7 \\
        NSGA-III                                    & .1062      &  .2320     & 1305.0 \\
        R-NSGA-III                                  & .1215      &  .2279     & 1304.3 \\
        U-NSGA-III                                  & .1314      &  .2567     & 1302.7 \\
        COMO-CMA-ES                                 & .0413      & 1.775      & 1141.1 \\
        MO-CMA-ES~\cite{nystrup2020hyperparameter}  & .1002      &  .0965     & 1317.4 \\
        $\mathcal{H}$:MO-CMA-ES*                    & .0912      & \tb{.0951} & \tb{1320.0} \\
        $\mathcal{H}$:NSGA-II*                      & .0999      &  .1141     & 1316.1 \\
    \bottomrule
    \end{tabular*}
    \smallskip
    \begin{tablenotes}
    \item[*] ParDen algorithm applied
    \end{tablenotes}
    \end{threeparttable}
\end{table}

\begin{figure*}[htbp]
\centering
\begin{tikzpicture}
    \begin{groupplot}[group style={
                      group name=myplot,
                      group size= 2 by 5},height=4.75cm,width=8.75cm]
        \nextgroupplot[ylabel={Return \%}, 
                      xmajorgrids=true,
                      ymajorgrids=true,
                      grid style={gray!30, loosely dashed},
                      legend entries={{Random 6510},{Grid Search~\cite{boyd2017multi}}},
                      legend style={at={(.9875,0.025)},anchor=south east},
                      legend cell align={left},
                      xticklabels={},
                      axis background/.style={fill=gray!7}]
                \addplot+[only marks,mark size = 1.75, mark options={
                    draw = black!90,
                    fill = black!70,
                    fill opacity=0.6,
                    draw opacity=0.3,
                 }]
                  table[
                  x=Risk,
                  y=Return,
                  col sep=comma,
                  ]{points_everything.csv};\label{plots:plot1}
                \addplot+[only marks,mark=square*, mark size = 1.75,mark options={
                    draw = Set1-B!90,
                    fill = Set1-B!70,
                }]
                  table[
                  x=Risk,
                  y=Return,
                  col sep=comma,
                  ]{points_boyd.csv};\label{plots:plot2}
        \nextgroupplot[%
                      xmajorgrids=true,
                      ymajorgrids=true,
                      legend entries={{Random 6510},{NSGA-II}},
                      legend style={at={(.9875,0.025)},anchor=south east},
                      legend cell align={left},
                      grid style={gray!30, loosely dashed},
                      xticklabels={},
                      yticklabels={},
                      axis background/.style={fill=gray!7}
                      ]
                \addplot+[only marks,mark size = 1.75, mark options={
                    draw = black!90,
                    fill = black!70,
                    fill opacity=0.6,
                    draw opacity=0.3,
                }]
                  table[
                  x=Risk,
                  y=Return,
                  col sep=comma,
                  ]{points_everything.csv};
                \addplot+[only marks,mark=diamond*, mark size = 1.75,mark options={
                    draw = Dark2-A!90,
                    fill = Dark2-A!70,
                }]
                  table[
                  x=Risk,
                  y=Return,
                  col sep=comma,
                  ]{points_nsga_ii.csv};\label{plots:plot3}
        \nextgroupplot[ylabel={Return \%}, 
                      xmajorgrids=true,
                      ymajorgrids=true,
                      legend entries={{Random 6510},{R-NSGA-II}},
                      legend style={at={(.9875,0.025)},anchor=south east},
                      legend cell align={left},
                      grid style={gray!30, loosely dashed},
                      xticklabels={},
                      axis background/.style={fill=gray!7}]
                \addplot+[only marks,mark size = 1.75, mark options={
                    draw = black!90,
                    fill = black!70,
                    fill opacity=0.6,
                    draw opacity=0.3,
                }]
                  table[
                  x=Risk,
                  y=Return,
                  col sep=comma,
                  ]{points_everything.csv};
                \addplot+[only marks,mark=triangle*, mark size = 1.75,mark options={
                    draw = Set1-A!90,
                    fill = Set1-A!70,
                }]
                  table[
                  x=Risk,
                  y=Return,
                  col sep=comma,
                  ]{points_r_nsga_ii.csv};\label{plots:plot4}
        \nextgroupplot[
                      xmajorgrids=true,
                      ymajorgrids=true,
                      legend entries={{Random 6510},{NSGA-III}},
                      legend style={at={(.9875,0.025)},anchor=south east},
                      legend cell align={left},
                      grid style={gray!30, loosely dashed},
                      xticklabels={},
                      yticklabels={},
                      axis background/.style={fill=gray!7}]
                 \addplot+[only marks,mark size = 1.75, mark options={
                    draw = black!90,
                    fill = black!70,
                    fill opacity=0.6,
                    draw opacity=0.3,
                 }]
                  table[
                  x=Risk,
                  y=Return,
                  col sep=comma,
                  ]{points_everything.csv};
                 \addplot+[only marks,mark=oplus*, mark size = 1.75,mark options={
                    draw = Dark2-B!90,
                    fill = Dark2-B!70,
                }]
                  table[
                  x=Risk,
                  y=Return,
                  col sep=comma,
                  ]{points_nsga_iii.csv};\label{plots:plot5}
        \nextgroupplot[ylabel={Return \%}, 
                      xmajorgrids=true,
                      ymajorgrids=true,
                      legend entries={{Random 6510},{R-NSGA-III}},
                      legend style={at={(.9875,0.025)},anchor=south east},
                      legend cell align={left},
                      grid style={gray!30, loosely dashed},
                      xticklabels={},
                      axis background/.style={fill=gray!7}]
                \addplot+[only marks,mark size = 1.75, mark options={
                    draw = black!90,
                    fill = black!70,
                    fill opacity=0.6,
                    draw opacity=0.3,
                }]
                  table[
                  x=Risk,
                  y=Return,
                  col sep=comma,
                  ]{points_everything.csv};
                \addplot+[only marks,mark=halfsquare left*, mark size = 1.75,mark options={
                    draw = Dark2-D!90,
                    fill = Dark2-D!70,
                }]
                  table[
                  x=Risk,
                  y=Return,
                  col sep=comma,
                  ]{points_r_nsga_iii.csv};\label{plots:plot6}
        \nextgroupplot[
                      xmajorgrids=true,
                      ymajorgrids=true,
                      legend entries={{Random 6510},{U-NSGA-III}},
                      legend style={at={(.9875,0.025)},anchor=south east},
                      legend cell align={left},
                      grid style={gray!30, loosely dashed},
                      xticklabels={},
                      yticklabels={},
                      axis background/.style={fill=gray!7}]
                  \addplot+[only marks,mark size = 1.75, mark options={
                     draw = black!90,
                     fill = black!70,
                     fill opacity=0.6,
                     draw opacity=0.3,
                  }]
                   table[
                   x=Risk,
                   y=Return,
                   col sep=comma,
                   ]{points_everything.csv};
                 \addplot+[only marks,mark=halfsquare right*, mark size = 1.75,mark options={
                     draw = Set1-B!90,
                     fill = Set1-B!70,
                 }]
                   table[
                   x=Risk,
                   y=Return,
                   col sep=comma,
                   ]{points_u_nsga_iii.csv};\label{plots:plot7}
        \nextgroupplot[ylabel={Return \%}, 
                      xmajorgrids=true,
                      ymajorgrids=true,
                      legend entries={{Random 6510},{COMO-CMA-ES}},
                      legend style={at={(.9875,0.025)},anchor=south east},
                      legend cell align={left},
                      grid style={gray!30, loosely dashed},
                      xticklabels={},
                      axis background/.style={fill=gray!7}]
                 \addplot+[only marks,mark size = 1.75, mark options={
                    draw = black!90,
                    fill = black!70,
                    fill opacity=0.6,
                    draw opacity=0.3,
                 }]
                  table[
                  x=Risk,
                  y=Return,
                  col sep=comma,
                  ]{points_everything.csv};
                \addplot+[only marks,mark=otimes*, mark size = 1.75,mark options={
                    draw = Dark2-F!90,
                    fill = Dark2-F!70,
                }]
                  table[
                  x=Risk,
                  y=Return,
                  col sep=comma,
                  ]{points_como_cma_es.csv};\label{plots:plot8}
        \nextgroupplot[
                      xmajorgrids=true,
                      ymajorgrids=true,
                      legend entries={{Random 6510},{MO-CMA-ES~\cite{nystrup2020hyperparameter}}},
                      legend style={at={(.9875,0.025)},anchor=south east},
                      legend cell align={left},
                      grid style={gray!30, loosely dashed},
                      xticklabels={},
                      yticklabels={},
                      axis background/.style={fill=gray!7}]
                 \addplot+[only marks,mark size = 1.75, mark options={
                    draw = black!90,
                    fill = black!70,
                    fill opacity=0.6,
                    draw opacity=0.3,
                 }]
                  table[
                  x=Risk,
                  y=Return,
                  col sep=comma,
                  ]{points_everything.csv};
                \addplot+[only marks,mark=halfcircle*, mark size = 1.75,mark options={
                    draw = Dark2-G!90,
                    fill = Dark2-G!70,
                }]
                  table[
                  x=Risk,
                  y=Return,
                  col sep=comma,
                  ]{points_mo_cma_es.csv};\label{plots:plot9}
        \nextgroupplot[xlabel={Risk \%}, 
                      ylabel={Return \%}, 
                      xmajorgrids=true,
                      ymajorgrids=true,
                      legend entries={{Random 6510},{$\mathcal{H}$:MO-CMA-ES}},
                      legend style={at={(.9875,0.025)},anchor=south east},
                      legend cell align={left},
                      grid style={gray!30, loosely dashed},
                      axis background/.style={fill=gray!7}
                      ]
                 \addplot+[only marks,mark size = 1.75, mark options={
                    draw = black!90,
                    fill = black!70,
                    fill opacity=0.6,
                    draw opacity=0.3,
                 }]
                  table[
                  x=Risk,
                  y=Return,
                  col sep=comma,
                  ]{points_everything.csv};
                \addplot+[only marks,mark=halfdiamond*, mark size = 1.75,mark options={
                    draw = Set1-A!90,
                    fill = Set1-A!70,
                }]
                  table[
                  x=Risk,
                  y=Return,
                  col sep=comma,
                  ]{points_mo_cma_es_surrogate.csv};\label{plots:plot10}
        \nextgroupplot[xlabel={Risk \%}, 
                      xmajorgrids=true,
                      ymajorgrids=true,
                      legend entries={{Random 6510},{$\mathcal{H}$:NSGA-II}},
                      legend style={at={(.9875,0.025)},anchor=south east},
                      legend cell align={left},
                      grid style={gray!30, loosely dashed},
                      yticklabels={},
                      axis background/.style={fill=gray!7}]
                 \addplot+[only marks,mark size = 1.75, mark options={
                    draw = black!90,
                    fill = black!70,
                    fill opacity=0.6,
                    draw opacity=0.3,
                 }]
                  table[
                  x=Risk,
                  y=Return,
                  col sep=comma,
                  ]{points_everything.csv};
                \addplot+[only marks,mark=pentagon*, mark size = 1.75,mark options={
                    draw = Set1-C!90,
                    fill = Set1-C!70,
                }]
                  table[
                  x=Risk,
                  y=Return,
                  col sep=comma,
                  ]{points_nsga_ii_surrogate.csv};\label{plots:plot11}
    \end{groupplot}
\end{tikzpicture}
\caption{Optimal Frontiers for each Method in Table \ref{tab:quality_indicators}}\label{fig:frontiers}
\end{figure*}
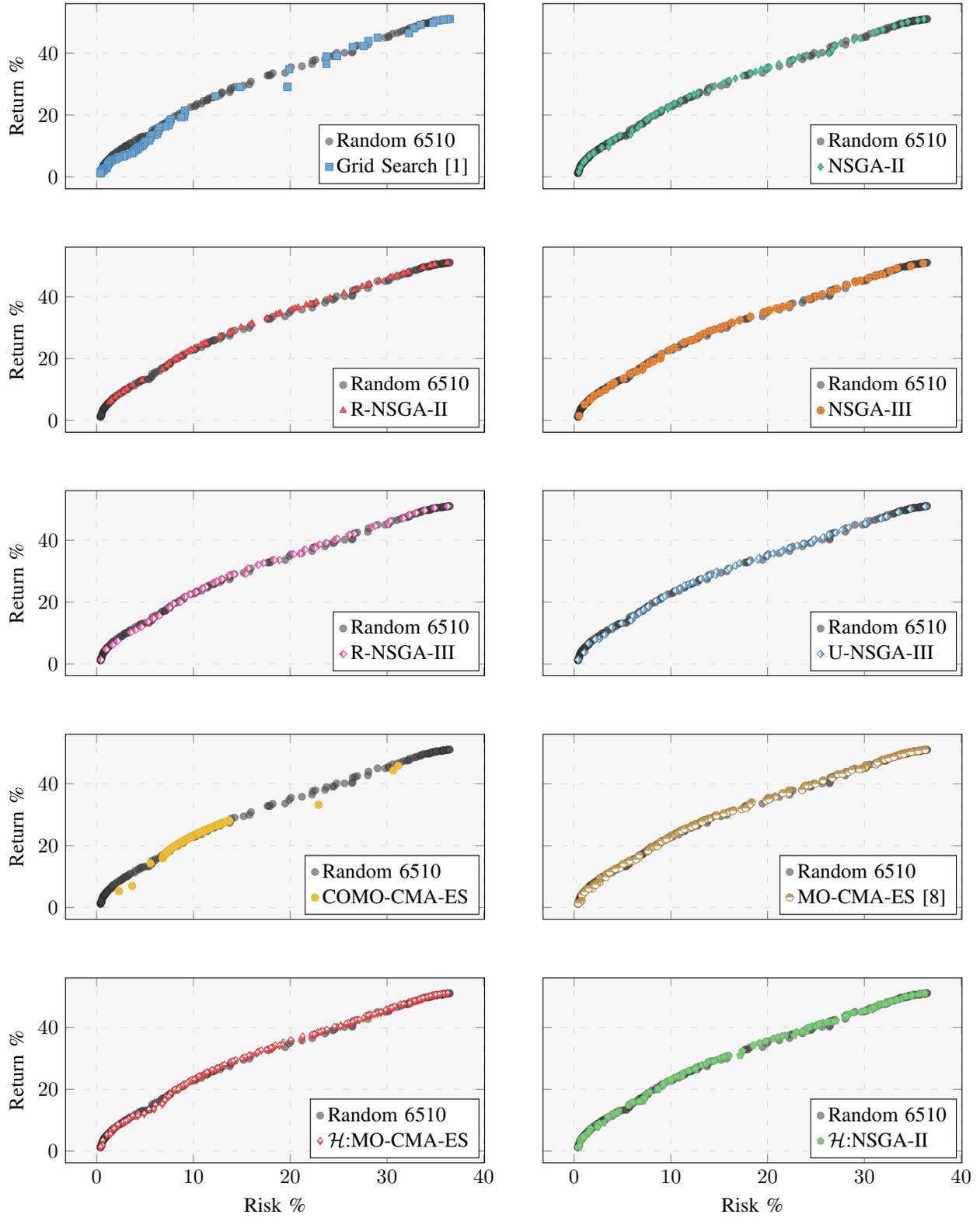

We note from Table~\ref{tab:perfermance_indicators} that all of the EAs perform similarly when comparing them using hypervolume. The difficulty with using hypervolume as a metric is that it is sensitive to the number of points in the Pareto front; thus, methods with a higher number of points have an advantage when compared via hypervolume. Next, we turn our attention to GD+ and IGD+ but are cognizant that neither of these is being measured with respect to the true Pareto front but instead measured against the Random $6510$ points as a reference. Fortunately, both GD+ and IGD+ are semi Pareto respecting, meaning they do not negatively impact the score of methods that outperform the reference Pareto front. In this respect, we note that although COMO-CMA-ES has one of the best GD+ scores, this has come at the cost of excessive crowing as seen in Figure \ref{fig:frontiers} and reinforced by the IGD+ score. Finally, we note that the surrogate assisted  $\mathcal{H}$:MO-CMA-ES has the best performing IGD+ score and the second-best GD+. Combined with the fact that it has the best HV, this indicates that our ParDen algorithm can outperform all other methods.

\begin{table}[htbp]
    \resizebox{1.\columnwidth}{!}{
    \begin{threeparttable}
    \caption{Quality Indicators}
    \label{tab:quality_indicators}
    \centering
   
    \begin{tabular}{lrrrrr}
    \toprule
        Method                                      & SR$@99$ & SR$@95$ & AESR$@99$ & AESR$@95$ & AGSR$@99$ \\
    \toprule
        Grid Search~\cite{boyd2017multi}            & 0.0     & 100.0   & -          & 510.0     & - \\
        NSGA-II                                     & 100.0   & 100.0   & 327.0      & \tb{63.0} & 9.9 \\
        R-NSGA-II                                   & 100.0   & 100.0   & 297.0      & \tb{63.0} & \tb{8.9} \\
        NSGA-III                                    & 100.0   & 100.0   & 405.0      & 81.0      & 12.5 \\
        R-NSGA-III                                  & 60.0    & 100.0   & 415.3      & 75.2      & 12.8 \\
        U-NSGA-III                                  & 70.0    & 100.0   & 338.6      & 96.0      & 10.3 \\
        COMO-CMA-ES                                 & 0.0     & 0.0     & -          & -         & - \\
        MO-CMA-ES~\cite{nystrup2020hyperparameter}  & 100.0   & 100.0   & 333.0      & 99.0      & 10.1 \\
        $\mathcal{H}$:MO-CMA-ES*                    & 100.0   & 100.0   & \tb{247.7} & 75.9      & 10.7 \\
        $\mathcal{H}$:NSGA-II*                      & 100.0   & 100.0   & 290.2      & 120.0     & 10.7 \\
    \bottomrule
    \end{tabular}
    \smallskip
    \begin{tablenotes}
    \item[*] ParDen algorithm applied
    \end{tablenotes}
    \end{threeparttable}
    }
\end{table}

\pgfplotsset{ every non boxed x axis/.append style={x axis line style=-}}
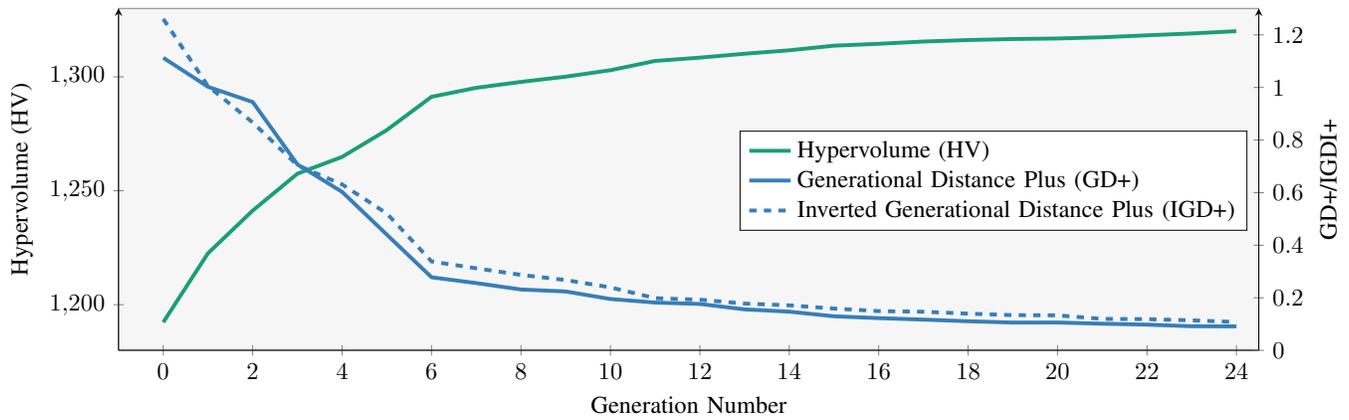
\begin{figure*}[htbp]
\centering
\resizebox{\textwidth}{!}{
\begin{tikzpicture}[scale=.75]
\begin{axis}[
    xmin  = -1,  xmax = 24.5,  
    ymin  =  1180, 
    ymax = 1330.0,
    ylabel=Hypervolume (HV),
    xlabel=Generation Number,
    axis lines = left,
    width=\textwidth,
    height=.9*\axisdefaultheight,
    axis background/.style={fill=gray!7, draw=gray},
    ]
\addplot [Dark2-A, ultra thick]
     table [x = it, y = val, col sep=comma] {convergence_hvs.csv};\label{hv}

\end{axis}

\begin{axis}[
    axis y line=right,
    axis x line=none,
    ymin = -0.0,
    ymax = 1.3,
    xmin = -1,
    xmax = 24.5,
    ylabel= GD+/IGDI+,
    legend cell align={left},
    legend style={at={(0.99,0.35)},
		anchor=south east,legend columns=1},
    width=\textwidth,
    height=.9*\axisdefaultheight
]

\addlegendimage{/pgfplots/refstyle=hv}\addlegendentry{Hypervolume (HV)}

\addplot [Set1-B, ultra thick]
table [ x = it, y = val, col sep=comma] {convergence_gdps.csv} ;\label{gpds}
\addlegendentry{Generational Distance Plus (GD+)}

\addplot [Set1-B, ultra thick, dashed]
table [ x = it, y = val, col sep=comma] {convergence_igdps.csv} ;\label{igpds}
\addlegendentry{Inverted Generational Distance Plus (IGD+)}

\end{axis}

\end{tikzpicture}
}
\caption{Convergence Plots of HV, GD+ and IGD+ Metrics for $\mathcal{H}$:MO-CMA-ES over Successive Generations}
\label{fig:convergence_plot}
\end{figure*}

\pgfplotsset{ every non boxed x axis/.append style={x axis line style=-stealth}}
\begin{figure*}
    \centering
    \resizebox{\textwidth}{!}{
\begin{tikzpicture}
\begin{axis}[
            xmin=-1.0,
            ymin=-1.0,
            xmax=38,
            ymax=55,
            xlabel=Risk \%,
            ylabel=Return \%,
            axis lines = left,
            width=\textwidth,
            height=.9*\axisdefaultheight,
            legend cell align={left},
            legend style={at={(0.99,0.05)},
		    anchor=south east,legend columns=1},
		    axis background/.style={fill=gray!7, draw=gray}
            ]
            
\addplot+[only marks, mark=halfdiamond*, color=Set1-A, opacity=1] table[col sep=comma,x={y22},y={x22}]{convergence_pfs.csv};\label{mo_advance}
 \addlegendimage{/pgfplots/refstyle=mo_advance}\addlegendentry{$\mathcal{H}$:MO-CMA-ES}
\pgfplotsinvokeforeach{0,...,22}{
  \addplot+[only marks, mark=halfdiamond*, color=Set1-A,opacity={0.1+0.01*#1}] table[col sep=comma,x={y#1},y={x#1}]
  {convergence_pfs.csv};
}

\end{axis}
\end{tikzpicture}
}
    \caption{Descending Opacity Illustrating the Advancing Optimal Frontier of $\mathcal{H}$:MO-CMA-ES from Generation to Generation}
    \label{fig:improving_pareto_plot}
\end{figure*}
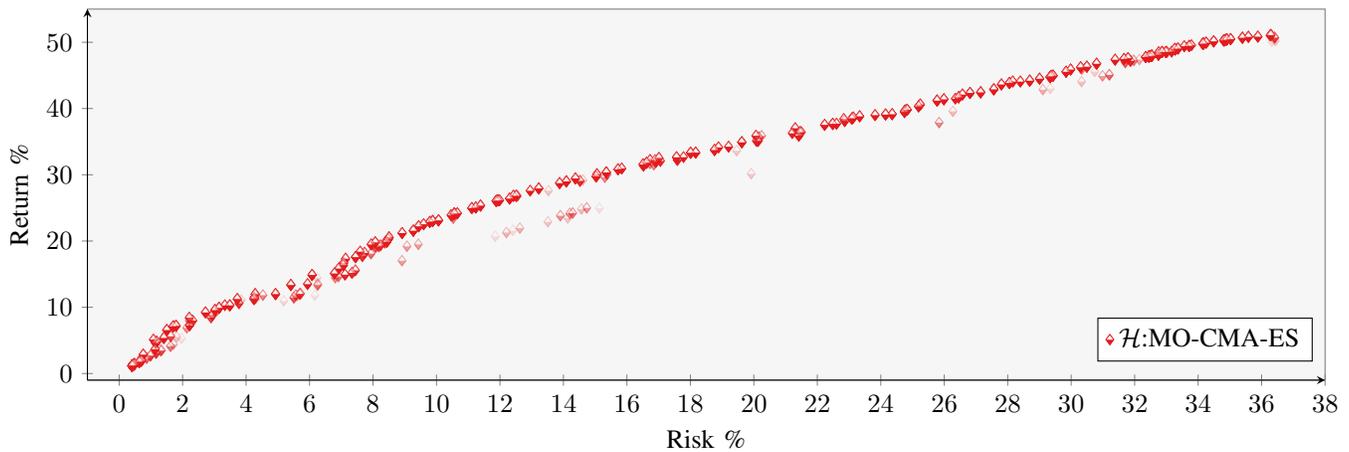

Although the ParDen algorithm can achieve superior performance with respect to HV, GD+, and IGD+, of interest is the quality of the algorithm in arriving at that solution. From Table~\ref{tab:quality_indicators}, we note that for both surrogate assisted methods, $\mathcal{H}$:MO-CMA-ES and $\mathcal{H}$:NSGA-II, the number of evaluations required to reach a solution within 99\% of random $6510$ reference HV is noticeably lower. In fact, $\mathcal{H}$:MO-CMA-ES requires almost one third or approximately $100$ fewer evaluations. Turing to Figure~\ref{fig:convergence_plot} and Figure~\ref{fig:improving_pareto_plot} we see the rapid initial gains in HV, GD+ and IGD+ for $\mathcal{H}$:MO-CMA-ES. Again, these results lend support to the efficacy of the ParDen algorithm, especially in cases where there is a realisable cost associated with each execution of the simulation. 

In previous work, Nydstrup \textit{et al.}~\cite{nystrup2020hyperparameter} mentions that one need only consider the number of evaluations. However, this is not entirely correct as EAs improve their solutions from generation to generation. As a result the total evaluations of the simulation must be considered with respect to incremental improvements between generations. To this end, we consider the number of generations required as a proxy for Efficiency. It is meaningful to make this assumption as all evaluations within a generation could be done in parallel and the time taken for an evaluation of the backtest is orders of magnitude larger than any of the operations performed by optimisation algorithms. Inspecting the Average number of Generations to a SR$@99$ (AGSR$@99$) we note that, although the surrogates assisted models can achieve better results with fewer evaluations of the simulation, they still require the same number of generations. This observation is significant since it highlights the need for further investigations into algorithms that are able to reduce the number of generations required to reach optimal solutions.

\section{Conclusion}\label{sec:conclusion}

In this paper, we have extended the existing research by applying a number of additional EA algorithms to the constrained multi-objective portfolio optimisation problem. We further built upon this work by introducing ParDen; a Surrogate assisted algorithm that, together with an NDScore, allows any generative or discriminative machine learning model to be used. Our results show the superiority of our ParDen framework across several performance and quality metrics. Our research highlights the need for further investigation into algorithms better suited to problems with computationally resource intensive objective evaluations. 

\bibliographystyle{IEEEtran}
\bibliography{main}

\begin{thebibliography}{10}
\providecommand{\url}[1]{#1}
\csname url@samestyle\endcsname
\providecommand{\newblock}{\relax}
\providecommand{\bibinfo}[2]{#2}
\providecommand{\BIBentrySTDinterwordspacing}{\spaceskip=0pt\relax}
\providecommand{\BIBentryALTinterwordstretchfactor}{4}
\providecommand{\BIBentryALTinterwordspacing}{\spaceskip=\fontdimen2\font plus
\BIBentryALTinterwordstretchfactor\fontdimen3\font minus
  \fontdimen4\font\relax}
\providecommand{\BIBforeignlanguage}[2]{{%
\expandafter\ifx\csname l@#1\endcsname\relax
\typeout{** WARNING: IEEEtran.bst: No hyphenation pattern has been}%
\typeout{** loaded for the language `#1'. Using the pattern for}%
\typeout{** the default language instead.}%
\else
\language=\csname l@#1\endcsname
\fi
#2}}
\providecommand{\BIBdecl}{\relax}
\BIBdecl

\bibitem{boyd2017multi}
S.~Boyd, E.~Busseti, S.~Diamond, R.~N. Kahn, K.~Koh, P.~Nystrup, J.~Speth
  \emph{et~al.}, ``Multi-period trading via convex optimization,''
  \emph{Foundations and Trends{\textregistered} in Optimization}, vol.~3,
  no.~1, pp. 1--76, 2017.

\bibitem{markowitz1952}
H.~Markowitz, ``Portfolio selection,'' \emph{The Journal of Finance}, vol.~7,
  no.~1, pp. 77--91, 1952.

\bibitem{art:doering2019metaheuristics}
J.~Doering, R.~Kizys, A.~A. Juan, {\`A}.~Fit{\'o}, and O.~Polat,
  ``Metaheuristics for rich portfolio optimisation and risk management: Current
  state and future trends,'' \emph{Operations Research Perspectives}, vol.~6,
  p. 100121, 2019.

\bibitem{art:loukeris2009numerical}
N.~Loukeris, D.~Donelly, A.~Khuman, and Y.~Peng, ``A numerical evaluation of
  meta-heuristic techniques in portfolio optimisation,'' \emph{Operational
  Research}, vol.~9, no.~1, pp. 81--103, 2009.

\bibitem{art:fernandez2015hybrid}
E.~Fernandez, C.~Gomez, G.~Rivera, and L.~Cruz-Reyes, ``Hybrid metaheuristic
  approach for handling many objectives and decisions on partial support in
  project portfolio optimisation,'' \emph{Information Sciences}, vol. 315, pp.
  102--122, 2015.

\bibitem{cvx}
M.~Grant and S.~Boyd, ``{CVX}: Matlab software for disciplined convex
  programming, version 2.1,'' \url{http://cvxr.com/cvx}, Mar. 2014.

\bibitem{grant2006disciplined}
M.~Grant, S.~Boyd, and Y.~Ye, ``Disciplined convex programming,'' in
  \emph{Global optimization}.\hskip 1em plus 0.5em minus 0.4em\relax Springer,
  2006, pp. 155--210.

\bibitem{nystrup2020hyperparameter}
P.~Nystrup, E.~Lindstr{\"o}m, and H.~Madsen, ``Hyperparameter optimization for
  portfolio selection,'' \emph{The Journal of Financial Data Science}, vol.~2,
  no.~3, pp. 40--54, 2020.

\bibitem{branke2009portfolio}
J.~Branke, B.~Scheckenbach, M.~Stein, K.~Deb, and H.~Schmeck, ``Portfolio
  optimization with an envelope-based multi-objective evolutionary algorithm,''
  \emph{European Journal of Operational Research}, vol. 199, no.~3, pp.
  684--693, 2009.

\bibitem{ruiz2010hybrid}
R.~Ruiz-Torrubiano and A.~Suarez, ``Hybrid approaches and dimensionality
  reduction for portfolio selection with cardinality constraints,'' \emph{IEEE
  Computational Intelligence Magazine}, vol.~5, no.~2, pp. 92--107, 2010.

\bibitem{ruiz2015memetic}
R.~Ruiz-Torrubiano and A.~Su{\'a}rez, ``A memetic algorithm for
  cardinality-constrained portfolio optimization with transaction costs,''
  \emph{Applied Soft Computing}, vol.~36, pp. 125--142, 2015.

\bibitem{baykasouglu2015grasp}
A.~Baykaso{\u{g}}lu, M.~G. Yunusoglu, and F.~B. {\"O}zsoydan, ``A grasp based
  solution approach to solve cardinality constrained portfolio optimization
  problems,'' \emph{Computers \& Industrial Engineering}, vol.~90, pp.
  339--351, 2015.

\bibitem{qi2017hybrid}
R.~Qi and G.~G. Yen, ``Hybrid bi-objective portfolio optimization with
  pre-selection strategy,'' \emph{Information Sciences}, vol. 417, pp.
  401--419, 2017.

\bibitem{book:eiben2015}
A.~Eiben and J.~Smith, \emph{Introduction to Evolutionary Computing}.\hskip 1em
  plus 0.5em minus 0.4em\relax Springer, 2015.

\bibitem{art:liu2020multi}
Q.~Liu, X.~Li, H.~Liu, and Z.~Guo, ``Multi-objective metaheuristics for
  discrete optimization problems: A review of the state-of-the-art,''
  \emph{Applied Soft Computing}, p. 106382, 2020.

\bibitem{art:deb2002fast}
K.~Deb, A.~Pratap, S.~Agarwal, and T.~Meyarivan, ``A fast and elitist
  multiobjective genetic algorithm: Nsga-ii,'' \emph{IEEE transactions on
  evolutionary computation}, vol.~6, no.~2, pp. 182--197, 2002.

\bibitem{art:zhang2007moea}
Q.~Zhang and H.~Li, ``Moea/d: A multiobjective evolutionary algorithm based on
  decomposition,'' \emph{IEEE Transactions on evolutionary computation},
  vol.~11, no.~6, pp. 712--731, 2007.

\bibitem{art:igel2007covariance}
C.~Igel, N.~Hansen, and S.~Roth, ``Covariance matrix adaptation for
  multi-objective optimization,'' \emph{Evolutionary computation}, vol.~15,
  no.~1, pp. 1--28, 2007.

\bibitem{art:helbig2014population}
M.~Helbig and A.~P. Engelbrecht, ``Population-based metaheuristics for
  continuous boundary-constrained dynamic multi-objective optimisation
  problems,'' \emph{Swarm and Evolutionary computation}, vol.~14, pp. 31--47,
  2014.

\bibitem{art:alsattar2020mogsabat}
H.~AlSattar, A.~Zaidan, B.~Zaidan, M.~A. Bakar, R.~Mohammed, O.~Albahri,
  M.~Alsalem, and A.~Albahri, ``Mogsabat: A metaheuristic hybrid algorithm for
  solving multi-objective optimisation problems,'' \emph{Neural Computing and
  Applications}, vol.~32, no.~8, pp. 3101--3115, 2020.

\bibitem{blank2020pymoo}
J.~Blank and K.~Deb, ``pymoo: Multi-objective optimization in python,''
  \emph{IEEE Access}, vol.~8, pp. 89\,497--89\,509, 2020.

\bibitem{art:ben2017universal}
M.~Ben~Salem, O.~Roustant, F.~Gamboa, and L.~Tomaso, ``Universal prediction
  distribution for surrogate models,'' \emph{SIAM/ASA Journal on Uncertainty
  Quantification}, vol.~5, no.~1, pp. 1086--1109, 2017.

\bibitem{art:zhou2006combining}
Z.~Zhou, Y.~S. Ong, P.~B. Nair, A.~J. Keane, and K.~Y. Lum, ``Combining global
  and local surrogate models to accelerate evolutionary optimization,''
  \emph{IEEE Transactions on Systems, Man, and Cybernetics, Part C
  (Applications and Reviews)}, vol.~37, no.~1, pp. 66--76, 2006.

\bibitem{art:wan2005simulation}
X.~Wan, J.~F. Pekny, and G.~V. Reklaitis, ``Simulation-based optimization with
  surrogate models—application to supply chain management,'' \emph{Computers
  \& chemical engineering}, vol.~29, no.~6, pp. 1317--1328, 2005.

\bibitem{art:wang2018random}
H.~Wang and Y.~Jin, ``A random forest-assisted evolutionary algorithm for
  data-driven constrained multiobjective combinatorial optimization of trauma
  systems,'' \emph{IEEE transactions on cybernetics}, vol.~50, no.~2, pp.
  536--549, 2018.

\bibitem{art:yang2019offline}
C.~Yang, J.~Ding, Y.~Jin, and T.~Chai, ``Offline data-driven multiobjective
  optimization: Knowledge transfer between surrogates and generation of final
  solutions,'' \emph{IEEE Transactions on Evolutionary Computation}, vol.~24,
  no.~3, pp. 409--423, 2019.

\bibitem{art:chugh2017data}
T.~Chugh, N.~Chakraborti, K.~Sindhya, and Y.~Jin, ``A data-driven
  surrogate-assisted evolutionary algorithm applied to a many-objective blast
  furnace optimization problem,'' \emph{Materials and Manufacturing Processes},
  vol.~32, no.~10, pp. 1172--1178, 2017.

\bibitem{art:stander2020data}
L.~Stander, M.~Woolway, and T.~van Zyl, ``Data-driven evolutionary optimisation
  for the design parameters of a chemical process: A case study,'' in
  \emph{2020 IEEE 23rd International Conference on Information Fusion
  (FUSION)}.\hskip 1em plus 0.5em minus 0.4em\relax IEEE, 2020, pp. 1--8.

\bibitem{book:selvi2018application}
S.~T. Selvi, S.~Baskar, and S.~Rajasekar, ``Application of evolutionary
  algorithm for multiobjective transformer design optimization,'' in
  \emph{Classical and Recent Aspects of Power System Optimization}.\hskip 1em
  plus 0.5em minus 0.4em\relax Elsevier, 2018, pp. 463--504.

\end{thebibliography}

\end{document}